\documentclass[preprint]{revtex4-1}
\usepackage[latin1]{inputenc}
\usepackage{graphicx}
\usepackage{amsmath}

\newcommand{\eps}{  {\varepsilon} }

\newcommand{\reali}{\mathinner{\bf R}}

\newcommand{\vett}[1]{ {\mathbf{ #1}} }
\newcommand{\udf}{ \stackrel{\mathrm{def}}{=} }
\newcommand{\tdot}[1]{\hskip2pt\ddot{\null}\hskip 3.2pt \dot{\null}\kern -5pt {#1}}
\newcommand{\xcm}{ {\vett x}_{cm}}
\newcommand{\dxcm}{ \dot{\vett x}_{cm}}
\newcommand{\ddxcm}{ \ddot{\vett x}_{cm}}

\newcommand{\rel}{ {\vett r}}

\newcommand{\dif}{ \,\mathrm{d}}

\begin{document}


\title{Classical Helium Atom with Radiation Reaction}

\author{G.~Camelio}

\affiliation{ Universit\`a degli Studi di Milano, Corso di Laurea in
 Fisica, Via Celoria  16,  I--20133 Milano (Italy)}
 \email{giovanni.camelio@gmail.com}
\author{A.~Carati}
 \affiliation{ Universit\`a degli Studi di Milano, Dipartimento di Matematica,
 Via Saldini 50,  I--20133 Milano (Italy)}
\email{andrea.carati@unimi.it}\email[]{ luigi.galgani@unimi.it}
\author{L.~Galgani}
 \affiliation{ Universit\`a degli Studi di Milano, Dipartimento di Matematica, Via Saldini 50,  I--20133 Milano (Italy)}
\email{luigi.galgani@unimi.it}
\date{\today}

\begin{abstract} 
We study a classical model of Helium atom in
which, in addition to the Coulomb forces,  the radiation reaction
forces  are taken into
account. This modification  brings   in the model  a new qualitative
feature of a global character. Indeed, as  pointed out  by
Dirac, in any model of classical electrodynamics of point particles
involving radiation reaction
one has  to eliminate, from the a priori  conceivable solutions of the
problem, those corresponding to the emission of an infinite amount of
energy. 
We show that the  Dirac prescription solves a problem of inconsistency
plaguing all available models which neglect radiation reaction, namely, the
fact that in all such models  most initial data
lead to a spontaneous breakdown of the atom.
A further modification is that the system thus acquires  a peculiar
form of dissipation. In particular, this  makes attractive an
invariant manifold of special physical interest, the zero--dipole
manifold, that corresponds to motions in which no energy is radiated away
(in the dipole approximation).
We finally study numerically the  invariant measure naturally  induced by
the time--evolution on such a manifold, and this corresponds to studying
the formation process of the atom.  Indications are given that such a
measure may be  singular with respect to that of Lebesgue.
\end{abstract}


\maketitle

\begin{quotation}
At variance with the existing works on
the dynamical properties of classical  atomic  systems (see for example Refs.
\onlinecite{nich,bohr,lang,kaneko1,kaneko2,poir,jayme,jayme2}), 
in this paper we take into 
account, for the case of the Helium atom,  the  effects on the
dynamics due to
the radiation emitted by the electrons during their motion. 
This introduces a qualitative change in the dynamics, with respect to the
purely Coulomb model,  because the  model now corresponds to a non
conservative dynamical system.  This leads to the appearing of an
``attractive'' manifold of stable periodics orbits,
 where attractive
is to be understood in a proper sense which will be explained below. The aim
of this paper is  to  study the properties of such a
manifold, through  analytical and  numerical
methods.
\end{quotation}

\section{Introduction}

The studies on the dynamical properties of classical atomic
models, in which only Coulomb forces are taken into
account, have a long history, going back to Nicholson, Bohr and
Langmuir (see Refs. \onlinecite{nich,bohr,lang}). The main result found in such
old works (see the work  \onlinecite{nich} of Nicholson  for the case of
Beryllium, at those times
called Nebulium) is that there exist periodic orbits having ``normal
mode frequencies'' whose ratios are near to those of the observed
spectral lines.  Among the recent works on this subject a
particularly relevant place is taken by the papers of De Luca
\cite{jayme,jayme2} on the Helium atom. This author 
went beyond the Coulomb approximation, taking
retardation of the electromagnetic forces into account. This was
obtained through a perturbation scheme which, truncated in a
suitable way, leads to the conservative Darwin lagrangian that one
finds illustrated in common textbooks\cite{landau,jackson}. 
In such a way the scale invariance of the purely
Coulomb model was  eliminated, and the result of Nicholson could be
improved, showing that for the Helium atom there exist periodic
orbits leading to dynamical frequencies which  themselves,
and not only their ratios, have a rather good agreement
with the empirical spectral frequencies.

On the other hand, it was  known since the time of Nicholson, and
confirmed  for example by Poirier \cite{poir},
that all such periodic orbits are unstable.
%
More recently  Yamamoto and Kaneko\cite{kaneko1,kaneko2}  showed, for
the case of Helium,  that the
instability of the periodic orbits  actually corresponds to a much more
general and acute form of instability. Indeed they  found that the
vast majority of initial data with negative energy lead
to the \emph{autoionization} of the atom, namely, to motions  in which one of the
electrons is  expelled. One can  see that this occurs also 
for a generic atom, and even in a stronger way, inasmuch as one meets
with motions in which all electrons
but one are expelled. 
So, the whole classical theory of atomic
models  seems to be plagued by a general failure,
for which no remedy is known.

%
One usually takes  the pragmatic attitude of just
forgetting the  autoionizing  motions. One should however 
indicate  an internal dynamical
justification of a more general character for such a selection of the 
initial data. Moreover, one might hope that such a general prescription
would  also  automatically explain why the system chooses to fall
(through some peculiar kind of dissipation) on the
physically interesting periodic orbits, for which an agreement
betweeen  dynamical and spectral frequencies is found.
Obviously, satisfying the latter point requires to
abandon a purely lagrangian, and thus conservative, 
approach to the problem, by  taking electromagnetic radiation into
account. 

This is precisely what we do  in  the present paper,
where we take into account  the emission of radiation  
due to the accelerations of the electrons  in the simplest
possible  way, namely, by adding to the Coulomb forces the radiation
reaction ones.  So, in  a sense our model 
is complementary to that of De Luca, who introduces only the
conservative terms produced by the Darwin approximation scheme, 
and neglects  dissipation. 
We leave for future work the study of a model which takes both
features into account.  

It will be shown here that the modification of the Coulomb
model that takes radiation into account, in the first place  gives a
model in which the autoionization problem no more arises. This is due
to the fact, first pointed out by Dirac in the general context of
classical electrodynamics of point particles\cite{dirac}, that
most initial data in the phase space suited to the model lead to
motions in which an infinite amount of energy is emitted, so that the
definition of the model has to be complemented by the explicit
prescription that such initial data have to be discarded. The
remaining  initial data (constituting a  set that will be called
here the \emph{Dirac \emph {or} physical manifold}) will be shown to
lead to motions in which the phenomenon of autoionization no more
occurs.
 So, the elimination of the
initial points in phase space leading to  autoionizing solutions appears no
more as a special trick to be strangely
introduced \emph{ad hoc}, but rather as a particular case
of a completely general prescription that, following Dirac, has always
to be introduced when dealing with classical models of
matter--radiation interaction involving  point particles.

Then we study the dynamics on the Dirac physical manifold, in which
the Dirac precription turns out to introduce a dissipation af a
peculiar type. We  show that the Dirac prscription
allows one to  find (and in a easy way) all periodic orbits, proving
furthermore that they form  in phase space a manifold. This is
just the \emph{zero--dipole manifold}, which is constituted of 
phase--space points leading to motions
that do not radiate energy away in the dipole approximation. A major aim
of this paper is to study such a manifold. We first   show that
it is an attractor, for initial states with negative energy. 
Then we study, by numerical methods, 
the invariant measure naturally
induced on it by the time--flow. This amounts to studying 
the \emph{formation process} of the Helium
atom,  namely, motions having initial data with the two
electrons coming from infinity that are then captured by the nucleus,
and finally fall on the zero--dipole manifold, 
where  emission of energy comes to an end. 

From the point of
view of the theory of dynamical systems, an interesting result
we find is  that, while  the attractor (the
zero--dipole manifold) is really  simple as a manifold 
(being just a portion of a linear subspace in the system's phase
space), it is the invariant measure induced on it by the time evolution
that is  very peculiar. Indeed it is presumably singular with respect to
the restriction of  the Lebesgue measure,  possibly having a fractal
structure. 

In Section~2 the model is introduced, and a preliminary analytical
discussion of its properties is given. In Section~3  the
numerical method for obtaining the invariant measure is described, 
and the numerical results are
presented. The conclusions follow.

\section{The model}
\subsection{The radiation reaction for a single point--charge}
Let us recall that, in the case of a single point--charge,
the emission of radiation is taken into account without introducing
the infinitely many degrees of freedom of the electromagnetic field,
through the expedient of adding
in the equation of motion for the particle   an ``effective
force''. This force, which is traditionally 
called   \emph{radiation reaction force}  and 
denoted by $\vett K^{rad}$,  is given by
\begin{equation}\label{kappa}
\vett K^{rad}= \frac 23\frac {e^2}{c^3} \tdot {\vett x}\ ,
\end{equation}
where $e$  is the charge  of the particle,  $c$ the speed of light, and
${\vett x}$ the position vector of the particle.
The procedure which leads to such  a force goes back to Planck, Lorentz
and Abraham, and was given a  final form in the work of Dirac \cite{dirac}
of  the year 1938, in which an extension to the relativistic
case was performed in an extremely elegant way. In its most
elementary form, which is illustrated in standard  textbooks (see
Refs.~\onlinecite{landau,jackson,pauli,heitler}), the
procedure amounts to requiring that the effective force $\vett K^{rad}$ 
produces an energy loss consistent with the power radiated away by an
accelerating particle according to the Larmor formula, namely,
 $(2/3)\, ({e^2}/{c^3})\, |\ddot{\vett  x}|^2$.
As is well known, this force can also be interpreted
  as being   produced by the regular part of the self--field of  the
  particle, the divergent part having been reabsorbed through mass
  renormalization.  

We illustrate now, in the simple  case of the free particle
in the nonrelativistic approximation, how  the
``runaway'' solutions then show up.  In terms of the particle's acceleration 
$\vett a=\ddot {\vett x}$,   the equation of motion for the free
particle  takes the form  
$$\eps \dot {\vett a} =\vett a\ ,
$$ 
with
\begin{equation}\label{epsilon}
\eps= \frac 23\, \frac {e^2}{mc^3} 
\end{equation}
($m$  being  the particle's mass), with solutions  
$\vett a(t)=\vett a_0\exp (t/\eps)$
depending parametrically on the initial acceleration $\vett a_0$. So,
for  generic initial data the solution  exponentially explodes, i.e., presents
\emph{runaway} character, a fact which makes no sense for a free particle. Thus
Dirac \cite{dirac}
introduces the explicit prescription that, for the case of a free
particle,  only the solutions with
vanishing initial acceleration, $\vett a_0=0$, be retained. Now,  the
phase space $\mathcal{P}$ mathematically suited to the considered
third--order equation of
motion, is the vector space $\reali^9$, a point of which is defined by
the coordinates  $(\vett x, \vett v, \vett a)$ of position, velocity
and acceleration. In such an ambient phase space $\mathcal{P}$, the
physically meaningful phase space is thus  the subset 
characterized by motions of nonrunaway type. This we will call
``\emph{Dirac \emph{or} physical  manifold}'', and coincides with
 the hyperplane $\vett a=0$. 

More in general, in the
presence of an external force  vanishing at infinity, one should  assume
that for scattering  states, in which the particle eventually behaves
as a free one, the
physical or nonrunaway manifold be defined by the asymptotic condition
\begin{equation}\label{condizione}
\ddot {\vett x}(t)\to 0\quad \mathrm{for}\quad t\to +\infty\ .
\end{equation}

Prescriptions of such a type may be formulated in a different way,
which captures another side of the problem. One makes reference to the
total energy emitted by the particle, namely, the quantity
\begin{equation}\label{radiated}
\Delta E= m\eps \int_0^{+\infty}|\ddot{\vett x}(t)|^2 \, d t\ ,
\end{equation}
and a physically natural requirement is then to restrict oneself to motions
for which the amount of energy radiated away during the whole motion
is finite, i.e., one has $\Delta E< \infty$.
For smooth motions, such as those we are considering which are
solutions of an ordinary differential equation, the latter condition
implies the asymptotic condition (\ref{condizione}). Thus 
condition (\ref{condizione}) 
turns out to be  justified also for motions not having
a  scattering character (as the bound states), for which the argument 
referring to the eventual free--type motion of the particle does not apply.

The asymptotic condition (\ref{condizione}) will be
referred to as the \emph{Dirac prescription}.
Due to the global character (with respect to time) of the latter,  
the dynamics on the Dirac physical manifold thus acquires 
quite peculiar fetures \cite{generale}; for example, 
it can be folded \cite{haag,cg}.

We will give arguments indicating that, for the Helium atom model with radiation
reaction, the analog of the  Dirac prescription automatically eliminates  the  
autoionization problem, and thus overcomes the main inconsistency 
of all classical  models which neglect radiation.

\subsection{Definition of the model} 

In the case of several particles one can operate as for
just one particle, by introducing a suitable radiation reaction force
acting on each particle, in such a way that the   power dissipated  by the whole
system be equal to that emitted in the dipole approximation 
(see \onlinecite{landau} sec. 9.2).
In the case of two particles the radiated power is  
$m\eps(\ddot {\vett  x}_1+\ddot {\vett   x}_2)^2$, where 
 $\vett x_1$, $\vett x_2$ are  the
position vectors of the particles (here, the position vectors
 of the two electrons with
respect to the nucleus, assumed to be fixed at the origin of  an inertial
frame). This gives   for the effective forces 
$\vett K_1^{rad}$ and $\vett K_2^{rad}$ acting on each electron the same expression,
namely,
$m\eps ( \tdot{\vett x}_1$ +$ \tdot{\vett x}_2 )$. Thus the model
is described by the system of equations 
\begin{align}\label{eq:1}
  \ddot {\vett x}_1 &=
  -\, \frac {2e^2}m\,  \frac{\vett x_1}{r^3_1} +\frac {e^2}m\,
\frac{\vett x_1-\vett x_2}{r^3_{12}}
  + \eps  \left( \tdot{\vett x}_1 + \tdot{\vett x}_2
    \right)\nonumber\\ 
    \ddot {\vett x}_2 &=
    -\,  \frac {2e^2}m\, \frac{\vett x_2}{r^3_2} +
\frac {e^2}m\, \frac{\vett x_2-\vett x_1}{r^3_{12}}
    +  \eps  \left( \tdot{\vett x}_2 + \tdot{\vett x}_1
      \right) 
\end{align}
(in the Gauss system, with $\eps$ given by (\ref{epsilon})).

One obtains the same expression for the radiative force acting on each
particle, also if
one considers  it as due to  the sum of the
(regular part of the) self--field,  and of the retarded field produced by the
other charge,  expanded with respect to the distance.
One can check this fact
through computations which are completely standard, though a little
cumbersome and not particularly illuminating. So we omit them.
The only point worth of mention is that the expansion in the
distance should be performed up to third order and not just to the second
one as  one finds in the textbooks; this is indeed the point which is
responsible for the appearing of the relevant nonconservative
term proportional to the third derivative of the center of mass. 

System (\ref{eq:1}), when 
expressed in terms of the center of mass ${\vett
  x}_{cm}$ and of the relative position vector $\vett r$, defined by
$$
{\vett x}_{cm}=(\vett x_1+\vett x_2)/2 \ , \quad\quad  \vett r= {\vett
  x}_2-{\vett x}_1\ ,
$$
takes the form 
\begin{align}\label{eq:2}
  \ddot {\vett r} &=  \frac {2e^2}m \,\frac{ \vett r }{r^3} +  \frac
        {2e^2}m \, 
\Big(
 \frac{\vett x_{cm} -\vett r/2}{\big|\vett x_{cm} -\vett r/2  \big|^3}
 - \frac{\vett x_{cm} +\vett r/2}{\big|\vett x_{cm} +\vett r/2
 \big|^3} \Big)\nonumber\\
 2\eps  \tdot{\vett x}_{cm} &=  \ddot {\vett x}_{cm}+ \frac{e^2}{m}\, \Big( 
 \frac{\vett x_{cm} -\vett r/2}{\big|\vett x_{cm} -\vett r/2  \big|^3}
 + \frac{\vett x_{cm} +\vett r/2}{\big|\vett x_{cm} +\vett r/2
 \big|^3} \Big) \ ,
\end{align}
in which there appears only one third--derivative
term, $\tdot{\vett x}_{cm}$, which enters just the equation for the
center of mass.
In conclusion, when radiation reaction is taken into account, the 
equations of motion in the variables ${\vett x}_{cm}$, $\vett r$ are
exactly the same as for   the purely Coulomb model, with the only  addition
of the radiation reaction force in the equation for the
center of mass.

So the ambient phase space $\mathcal {P}$ suited for the motions of
the two point
electrons is  the vector space $\reali^{15}$, a point of which is
defined by the coordinates 
$$
{\vett x}_{cm}\ ,\ \dot{\vett x}_{cm}\ ,\ \ddot{\vett x}_{cm} \ , 
\ \vett r \ , \ \dot{\vett r}\ ,
$$
Now, the radiation reaction term acts in the equation for the center
of mass just  as in the case of  a single particle.
Thus, in the spirit of the classical work of Dirac, we will define our model 
by complementing the system of equations (\ref{eq:1}) or (\ref{eq:2}) through
the Dirac prescription: \emph{in the ambient phase space
  $\mathcal{P}=\reali^{15}$
the only admitted points  are those leading to motions that satisfy
 the asymptotic condition}  
 \begin{equation}\label{eq:3}
\ddot{\vett x}_{cm} (t) \to 0 \qquad \mbox{for}\quad   t \to +\infty \ .
\end{equation}

As in the case of the single particle, this   corresponds to the
physical condition that the
amount of energy radiated away during the whole motion  be finite. 
This prescription implicitly introduces a
selection among the allowed initial data, and the subset of  the allowed
points  in the ambient phase space $\mathcal P$
 will be called the \emph{Dirac \emph{or} physical manifold}.  We will show that 
 this restriction  eliminates the autoionization problem, and 
 makes the zero--dipole manifold become attractive for initial data of
 negative energy.

\subsection{Simple analytical deductions}
One immediately sees
that there exists a particular invariant submanifold of the   Dirac  physical
 manifold, which will play a fundamental role in this
work. It is the \emph
{zero--dipole manifold}, defined by the condition  $\xcm(t) =0$ for
all times, which corresponds to the hyperplane
\begin{equation}\label{eq:4}
\xcm = 0 \ , \quad \dxcm = 0 \ , \quad  \ddxcm = 0 \ .
\end{equation}
Invariance is immediately checked.  One also easily checks
that such a manifold is composed of orbits which are solutions of the
purely  Coulomb model in the unknown $\rel(t)$ with a suitable
charge, and are thus periodic in the case of negative energy. 

We will make use of  an energy theorem. This is 
immediately established by multiplying, 
as usual, equations \eqref{eq:1} by $\dot {\vett x}_1$ and $\dot
{\vett x}_2$ respectively, 
adding them and using  the Leibniz formula for the product
$\tdot{\vett x}_{cm}\dot{\vett x}_{cm}$. This gives
$$
\frac{\mathrm{d}~}{\mathrm{d}t}\left( 
T + V -4 m\eps \, \ddxcm \cdot \dxcm \right ) = - 4m\eps \, |\ddxcm|^2 \le 0 \ ,
$$
where $T$ is the kinetic energy, $V$ the potential
energy corresponding to the  Coulomb forces, and $4m\eps\, \ddxcm \cdot\dxcm$
the so--called Schott term. So, the quantity 
$$
\mathcal{E} = T + V - 4m\eps \ddxcm \cdot \dxcm 
$$ 
may be simply called  the \emph{energy}
of the system, (while the quantity
$E=T+V$ may  be called the \emph{mechanical energy}), and $\mathcal{ E}$
turns out to be a non increasing function of time. Obviously
the two energies coincide, $\mathcal{E}=E$,  on the
zero--dipole manifold.

From the energy theorem one immediately gets
 that periodic orbits necessarily belong to the
zero--dipole manifold. Indeed, just integrating the energy
equation over a period, a periodic orbit is seen to necessarily have
vanishing center of mass acceleration, and consequently is seen to belong to the
zero--dipole manifold. In addition, the portion
of the zero--dipole manifold with negative energy is wholly foliated 
by periodic orbits which, as already mentioned, are just the periodic
orbits for the two--body Coulomb problem with a suitable charge.

Using the  energy theorem one can also make more precise the
notion of an autoionizing motion. We recall that the instability  problem 
that plagues the purely
Coulomb model is that the vast majority of initial states with
negative  energy lead to motions in which one of the
electrons is expelled to infinity, so that  the atom is unstable. We
show instead that the analogous situation does not occur in the model
with radiation reaction, for initial states with negative
 energy. 

Indeed, if autoionization occurs with the
two electrons both escaping to infinity, then both the potential and
the Schott term  finally vanish, so that the  
energy  finally
becomes positive, i.e. it  \emph{has increased}, against the
energy theorem. If instead autoionization occurs with only one
electron, say the first one, escaping to infinity (in a nonrunaway
fashion), then the equation of motion for $\vett x_2$ reduces to
\begin{equation}\label{singolo} 
\ddot{\vett x}_2 = -\, \frac {2e^2}m \, \frac{\vett x_2}{r_2^3} + 
\eps\tdot{\vett x}_2\ ,
\end{equation}
which is the equation for just one electron with radiation
reaction, in the external field of the nucleus. On the other hand, as shown in
papers \onlinecite{carati} and \onlinecite{marino}, starting from initial data 
with negative  energy,  equation (\ref{singolo}) 
admits only runaway solutions.

 Furthermore, the zero--dipole manifold is   an attractor 
for initial states  with negative  energy, as a consequence
 of the asymptotic Dirac condition (\ref{eq:3}). This can be seen
 by rewriting the equation for the center of mass 
in the  integro--differential form
$$
\ddot{\vett x}_{cm}(t)=-\, \frac
1{e^{-t/2\eps}}\int_t^{+\infty}e^{-s/2\eps}\, f\big(\vett x_{cm}(s),
\vett r(s)\big)\, d\, s
$$
where $f$ is the function defined by the last two terms at the
right hand side of the second equation in (\ref{eq:2}).  So, from the  de
L'Hopital rule the property $f \to 0$ for $t\to +\infty$
follows. Finally, the result $\vett x_{cm}(t)\to 0$ for $t\to +\infty$
follows from the form of the function $f$, using the previously proven 
fact   that for negative energies the motion of the center of mass is bounded.  

In conclusion,
following the Dirac prescriprion of restricting the ambient phase
space to the Dirac physical   manifold, 
on the one hand the difficulty of the generic autoionization is
eliminated, and on the other hand 
our dynamical
system presents a \emph{dissipative} character, because all initial states
with negative energy are definitively attracted to the invariant zero--dipole
manifold, corresponding to motions that do not radiate energy away
(in the dipole approximation). However, in the next
section we will show  that the electrodynamical  dissipation due 
to radiation reaction
has a peculiar character with respect to that of the familiar dissipative systems
such as the  Lorenz one.

\section{The natural invariant measure on the attractor: numerical results}



It is simple to check that the Lebesgue measure in phase space,
 restricted to the
zero--dipole manifold, is  invariant under the flow. However,
other, actually infinitely many, invariant measures exist,  this being
 a general property of any 
continuous group $\Phi^t$ of  diffeomorfisms on a smooth
manifold $\mathcal{M}$. Indeed,  following Krylov and Bogolyubov \cite{krybog}, for
any given measure $\mu_0$ with $\mu_0(\mathcal{M})<\infty$, a corresponding
invariant measure $ \mu$ can be constructed by defining 
\begin{equation}\label{tilde}
 \mu(B) = \lim_{t\to+\infty} \frac  1t \int_0^t
\mu_0(\Phi^{-s}B) \dif s \ .
\end{equation}
The intuitive meaning of the measure $ \mu$
is the fraction of the initial points of phase space 
(chosen according the given  probability $\mu_0$)
which fall on any given set $B$. So, if we choose  for  $\mu_0$ the restriction of 
the Lebesgue measure to  the Dirac physical  manifold, and  take
initial data with   the two electrons coming freely from  infinity, 
then from the physical point of view   
the measure $\mu$ will  describe the state of the atom at
the end of the formation process. 

%
Our aim is now to find a numerical scheme for
constructing  the measure $\mu$. For the sake of simplicity, in
the implementation
we will restrict ourselves to the case of
planar motions. So, the  phase space corresponding to equations \eqref{eq:2}
has dimension 10, and the  Dirac physical  manifold has at most
dimension 8, while the zero--dipole manifold has dimension 4. 

It goes without saying that the motions are to be obtained by
integrating the system backward in time. Indeed
the Dirac physical manifold is defined only implicitly, through the 
asymptotic condition (\ref{eq:3}). Thus  one concretely has to operate 
in the ambient phase space $\mathcal P$, in which the Dirac manifold has vanishing
measure. Besides, even if one were able to take initial data on the
Dirac manifold, by integrating forward in time numerical errors
would make the orbits escape from it, and in a runaway fashion. 
Such problems are overcome by the standard
procedure of integrating the equations of motion  backward in time. In
such a way one is guaranteed to approach (in a time of the order
$2\eps$) the Dirac manifold. By the way, the integration
beackward in time is just the one needed in order to compute the
asymptotic measure $\mu$ according to the definition \eqref{tilde}.


From the numerical point of view, the only practical way in
which a measure can be described, is to compute its density
$\rho(\vett x)$. Here we denote  by $\vett x$ the coordinates of a
point in the  ambient phase space $\mathcal {P}$.
So we implemented the following algorithm. Taking $\vett x$ on the
zero--dipole manifold as  initial datum, we follow, backward
in time, the time
evolution  of a small hypercube of equal sides $|\delta \vett
x_j|$ ($j=1,\ldots,10$), having $\vett x$ as one of its
vertices, i.e., we follow the evolution of the points $\vett x_{j}=
\vett x +\delta \vett x_j $, and define, as usual, the side of the evolved
hypercube by $\delta \vett x_j (t)=\Phi^t \vett x_{j}- \Phi^t\vett x$. 
Notice that the hypercube does not belong to the zero--dipole manifold
(not even to the Dirac one); however,
in our specific case, the integration of the equations of motion,
backward in time, insures that the trajectories fall on the Dirac
manifold after a very short transient (of the order $2\eps$), 
because the runaway component
of the motion is  exponentially damped as one proceeds backward
in time. So, the hypercube becomes squeezed on the physical manifold,
and one of its 8--dimensional faces becomes tangent to the manifold. To get
the density $\rho(\vett x)$, it is then 
sufficient to compute the area of the latter face, 
divided by the initial area, and then calculate its 
time--average  along
the trajectory of $\vett x$ (which obviously belongs to the 
zero--dipole manifold for all times).

We  chose to numerically integrate the system with a fourth order
Runge--Kutta method and autoadaptation of the time grid.  
We take units with $m = 1$ and  $e =1$,
completing them by taking as unit of length
the Bohr radius $R_B$, as usually done in atomic physics. In this way
$\eps$ would take the value $1/(137)^3$. However, 
as we are only interested in the qualitative behaviour of the system,
in order to reduce the computational time 
we decided to take instead $\eps=10^{-2}$. 
The autoadaptation of the time grid was
implemented by choosing the integration step $h$ according to
$$
h = h_0 \cdot \max\, \big(\min (r_1 , r_2 ,r_{12} , r_{max} ),\,  r_{min} \big) \ ,
$$
where $h_0 =10^{-4}$, while
$r_1$ , $r_2$ and $r_{12}$ are the distances of the first
and second electron from the nucleus and their mutual distance
respectively, while $r_{min} = 0.001$ and $r_{max} = 1$ are factors 
introduced to avoid
that the integration step becomes too small or too large. The sides $|\delta \vett
x_j|$ of the initial hypercubes were all taken equal to $10^{-4}$.

Once the trajectories are known (numerically), there remains the
problem  of how to identify  the face of the hypercube lying on the
Dirac manifold (which is utterly unknown). In  order to solve this problem,
one can start from the fact that all sides  $\delta \vett x_j $
are essentially contained in the tangent plane to the manifold. Then
this plane can be found by determining
the 8--dimensional hyperplane to which the vectors $\delta \vett x_j $ are closer.
As an hyperplane
is defined by two independent unit vectors orthogonal to it,  say
$\vett a_1$ and $\vett a_2$,
then the tangent plane will be defined by the unit vectors which minimize 
$\sum_j\Big[ (\vett a_1\cdot\delta\vett x_j)^2 + (\vett a_2\cdot\delta\vett
x_j)^2 \Big]$, i. e.,  the sum of the distances of the points $\delta \vett x_j $
from the plane. In  other terms one has to minimize the function
$$
S(\vett a_1,\vett a_2)\udf \vett a_1^T B \vett a_1 + \vett a_2^T B \vett a_2
$$ 
where the symmetric matrix $B$ is defined by 
$$
B=[\delta \vett x_1,\ldots,\delta \vett x_{10}]^T\cdot [\delta \vett
  x_1,\ldots,\delta \vett x_{10}] \ .
$$
As  is well known, the two independent unit vectors which minimize $S$  are the 
eigenvectors corresponding to the two smallest eigenvalues
$\lambda_1$, $\lambda_2$ of $B$. 
With a little reflection, one also gets that the
area of the tangent face is nothing but the square root
$\sqrt{\lambda_3 \ldots\lambda_{10}}$ of the product
of the remaining 8 eigenvalues $\lambda_i$, $i=3,\ldots,10$ 
of the matrix $B$.

In  summary, the numerical procedure consists in taking for
$\rho(\vett x)$ the time--average
of $\sqrt{\lambda_3\ldots\lambda_{10}}$. 
Notice that,  as occurs in the computation of the Lyapunov exponents, 
the sides $\delta
{\vett x}_j(t)$ can grow too much during the evolution. 
If, for some $j$, $|\delta \vett x_j|$ becomes larger than $10^{-3}$, we
renormalize the cube choosing a new one with sides directed as
the eigenvectors of $B$ and again of size $10^{-4}$, and the  average
is then accordingly modified.

Just to give an indication of the form of the natural  measure,  
in the zero--dipole manifold   we chose $10^5$ initial points $\vett x$ 
uniformly distributed in the cube 
$$
 \left\{
 \begin{array}{rcl}
   \vett r & \in & [0.5,4.0]\times[0.5,4.0] \\
   \dot{\vett r} & \in & [-1.5,1.5]\times[-1.5,1.5] \ ,
 \end{array}
 \right. 
$$
with the condition that they have negative energy.
 We  checked that 
 averaging  over a period is enough to insure that the time
average is stabilized.
\begin{figure}
    \includegraphics[width=0.5\textwidth]{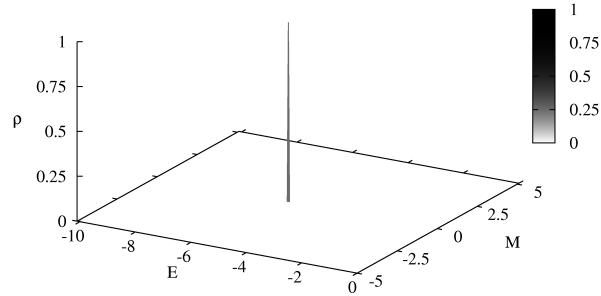}
  \caption{\label{fig:1} Invariant density $ \rho$ on the invariant
   zero--dipole manifold, projected on a rectangle of the ($E,M$) plane, of
   energy  and angular momentum.}
\end{figure}

\begin{figure}
   \includegraphics[width=0.5\textwidth]{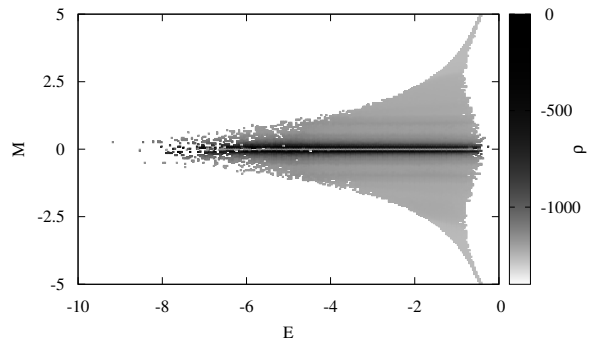}
  \caption{\label{fig:2}Same as Figure~\ref{fig:1}, 
 with the values of $ \rho$  represented in a gray decimal logarithmic
 scale, covering a thousand orders of magnitude.}
\end{figure}

\begin{figure}
    \includegraphics[width=0.4\textwidth]{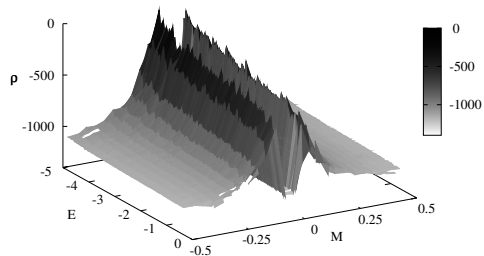}
  \caption{\label{fig:3} Same as Figure~\ref{fig:1}, restricted to a
    smaller portion of the original rectangle,  with  the ordinates
    in a  decimal logarithmic scale as in Figure~\ref{fig:2} .}
\end{figure}

Now, the zero--dipole manifold is 4--dimensional, so that in principle
$\rho$ depends on four variables (i.e., position and velocity).  
Recall however  that on such a manifold 
the motion is a purely Coulomb one, so that it is integrable. Then,
thinking in terms of action--angle variables, the
interesting ones  are just the two  actions, which in our
case are  $E^{-1/2}$,  $\ E$ being the (mechanical) energy, 
and the angular momentum $M$,
whereas  the  angles, i.e., the orientation of the orbit and the starting
point of the motion on the orbit, are irrelevant.
So we chose to integrate away the two angles 
from the density $\rho$, and to analyze the corresponding  reduced
density, $\rho(E,M)$ (by abuse of language still denoting it  by
$\rho$).

The result is shown in Figures~\ref{fig:1}--\ref{fig:3}.
In Figure~\ref{fig:1} the reduced density $\rho$ is reported versus
 $E$ and $M$. While a priori the measure   could be spread over the whole
rectangle, the figure  shows that it  is actually concentrated in an
extremely small area.  This is more clearly exhibited in
Figure~\ref{fig:2}, which still covers the whole rectangle, but gives
the density in a decimal lograthmic gray scale, which covers a thousand
orders of magnitude.
Figure~\ref{fig:3}  is instead the same as Figure~\ref{fig:1},
restricted to a smaller portion of the rectangle, still with the ordinates in
a decimal logarithmic scale. It shows  that the 
 function $\rho (E,M)$ has a very complex structure, with
values  ranging  over a thousand  orders of magnitude in an apparently  nonsmooth
way. This fact  might suggest  that the measure is 
 not absolutely continuous with respect to the restriction of the Lebesgue
one, possibly having some kind of fractal structure, but we leave this
problem for  possible future work.

In any case, we have shown that the peculiar character of dissipation
entering electrodynamics of point particles restricted to the
Dirac physical manifold, makes the final dynamics 
quite different from that of more familiar dissipative systems such as the
Lorenz one. Indeed in the latter case
it is the invariant
attractor  that has a \emph{strange} 
character, while the natural  invariant measure
induced on  by the dynamics  is somehow trivial, as corresponding 
to that of a hyperbolic system. In the former case, instead,
the invariant manifold is trivial (being just the hyperplane 
${\vett x}_{cm} = 0, \dot{\vett x}_{cm} = 0, \ddot{\vett x}_{cm} = 0$ 
in the ambient phase space), while  it is the
invariant measure which apparently has a strange character.

\section{Conclusions}
We investigated the main modifications that are introduced in
the Coulomb  Helium atom model when the radiation reaction forces are taken into
account in the simplest possible way, and the asymptotic  Dirac
prescription is consequently introduced, according to which   
the phase space has to be restricted to the submanifold of points
leading to   motions with a finite emitted energy.

The first qualitative result we found is that the problem of the 
breakdown of the atom  (corresponding to  autoionization for generic
initial data with negative energy), which makes the Coulomb model 
 inconsistent,
is now  eliminated. The second one  is the existence of an invariant 
manifold of stable periodic orbits, that moreover is attractive. 
Such a manifold is the zero--dipole one, on which the system does not
radiate energy away, in the dipole approximation.

Finally,    the invariant measure naturally  
induced by the time--flow
on the zero--dipole manifold was studied numerically. We showed that such a measure 
 is far from
trivial, being presumably   non absolutely continuous with respect to
the restriction of the Lebesgue measure, possibly with some kind of  
fractal structure.
This is at variance with the known examples of dissipative systems,
for which the measure is trivial, while it is the attractor that is strange
(i.e. a fractal set). This is apparently  a consequence of the peculiar
character of the radiation reaction force, which makes the system dissipative 
when restricted to the Dirac physical manifold, while making it expansive
in the ambient phase space.

A very interesting problem that remains open
is that of comparing  the special   periodic orbits which are  
 selected on the zero--dipole manifold as the most probable ones
according to the natural measure, 
with those, apparently of special physical relevance,
 studied by De~Luca.
We leave this point for  future work.

\begin{acknowledgements}
The first  author (G.C.) thanks  the Malegori family for  supporting
his studies at the Corso di Laurea in Fisica of the Milan University
through a ``Franca Erba scholarship''. 
\end{acknowledgements}

\end{document}